\documentstyle[mprocl]{article}

\newcommand{\beq}[1]{\begin{equation}\label{#1}}
\newcommand{\eeq}{\end{equation}}
\newcommand{\bear}[1]{\begin{eqnarray}\label{#1}}
\newcommand{\ear}{\end{eqnarray}}

\def\be{\begin{equation}}
\def\ee{\end{equation}}
\def\bea{\begin{eqnarray}}
\def\eea{\end{eqnarray}}

\newcommand{\nn}{\nonumber}

\newcommand{\N}{ \mbox{\rm I$\!$N} }
\newcommand{\R}{ \mbox{\rm I$\!$R} }

\newcommand{\ba}{\begin{eqnarray}}
\newcommand{\ea}{\end{eqnarray}}

\newcommand{\sign}{ \mbox{\rm sign} }
\newcommand{\e}{ \mbox{\rm e} }
\newcommand{\eps}{ \varepsilon }

\newcommand{\p}{\partial}

\newcommand{\tri}{\Delta}
\newcommand{\sq}[1]{\sqrt{|#1|}}

\begin{document}

\title{MULTIDIMENSIONAL 
GRAVITY WITH P-BRANES } 

\author{V.D.IVASHCHUK and V.N. MELNIKOV }

\address{
Center for Gravitation and Fundamental Metrology, VNIIMS,
\\ 3-1 M. Ulyanovoy Str., Moscow, 117313, Russia}

\maketitle\abstracts{ }


\section{The model}
\setcounter{equation}{0}

We consider the model governed by the action
\bear{2.1i}
S =&& 
\int_{M} d^{D}z \sqrt{|g|} \{ {R}[g] - 2 \Lambda - h_{\alpha\beta}\;
g^{MN} \partial_{M} \varphi^\alpha \partial_{N} \varphi^\beta
\\ \nn
&& - \sum_{a \in \Delta}
\frac{\theta_a}{n_a!} \exp[ 2 \lambda_{a} (\varphi) ] (F^a)^2_g \},
\ear
where $g = g_{MN} dz^{M} \otimes dz^{N}$ is the metric,
$\varphi=(\varphi^\alpha) \in \R^l$
is a vector from dilatonic scalar fields,
$(h_{\alpha\beta})$ is a non-degenerate $l\times l$ matrix ($l\in \N$),
$\theta_a = \pm 1$, $F^a =  dA^a$
is a $n_a$-form ($n_a \geq 1$) on a $D$-dimensional manifold $M$,
$\Lambda$ is cosmological constant and $\lambda_{a}$ is a $1$-form
on $\R^l$: $\lambda_{a} (\varphi) =\lambda_{a \alpha} \varphi^\alpha$,
$a \in \Delta$, $\alpha=1,\ldots,l$. In (\ref{2.1i}) 
$\Delta$ is some finite set. In the models
with one time all $\theta_a =  1$  when the signature of the metric
is $(-1,+1, \ldots, +1)$.

We consider the manifold
\beq{2.10g}
M = M_{0} \times \ldots \times M_{n}
\eeq
with the metric
\beq{2.11g}
g= e^{2{\gamma}(x)} g^0  +
\sum_{i=1}^{n} e^{2\phi^i(x)} g^i ,
\eeq
where $g^0  = g^0 _{\mu \nu}(x) dx^{\mu} \otimes dx^{\nu}$
is an arbitrary metric with any signature on the manifold $M_{0}$
and $g^i  = g^{i}_{m_{i} n_{i}}(y_i) dy_i^{m_{i}} \otimes dy_i^{n_{i}}$
is a metric on $M_{i}$  satisfying the equation
\beq{2.13}
R_{m_{i}n_{i}}[g^i ] = \xi_{i} g^i_{m_{i}n_{i}},
\eeq
$m_{i},n_{i}=1, \ldots, d_{i}$; $\xi_{i}= {\rm const}$,
(thus, $(M_i, g^i )$  is Einstein space)
$i=1,\ldots,n$. 
The functions $\gamma, \phi^{i} : M_0 \rightarrow \R$ are smooth.

Each manifold $M_i$ is assumed to be oriented and connected,
$i = 1,\ldots,n$. Then the volume $d_i$-form
$\tau_i  = \sqrt{|g^i(y_i)|}
\ dy_i^{1} \wedge \ldots \wedge dy_i^{d_i}$
and the signature parameter
$\eps(i)  = \sign \det (g^i_{m_{i}n_{i}}) = \pm 1$
are correctly defined for all $i=1,\ldots,n$.

Let
$\Omega_0$
be a set of all subsets of
$I_0\equiv\{ 1, \ldots, n \}$.
For any $I = \{ i_1, \ldots, i_k \} \in \Omega_0$, $i_1 < \ldots < i_k$,
we define a form
\beq{2.17i}
\tau(I) \equiv \tau_{i_1}  \wedge \ldots \wedge \tau_{i_k},
\eeq
of rank  $d(I)  = d_{i_1} + \ldots + d_{i_k}$,
and a corresponding $p$-brane submanifold 
$M_{I} \equiv M_{i_1}  \times  \ldots \times M_{i_k}$,
where $p=d(I)-1$.
We also define $\eps$-symbol 
$\eps(I) \equiv  \eps(i_1) \ldots \eps(i_k)$.
For $I = \emptyset$ we put  $\tau(\emptyset) = \eps(\emptyset) = 1$,
 $d(\emptyset) = 0$.

For fields of forms we adopt the following "composite electro-magnetic"
ansatz
\beq{2.27n}
F^a=\sum_{I\in\Omega_{a,e}}F^{(a,e,I)}+\sum_{J\in\Omega_{a,m}}F^{(a,m,J)},
\eeq
where
\bear{2.28n}
F^{(a,e,I)}=d\Phi^{(a,e,I)}\wedge\tau(I), \\ \label{2.29n}
F^{(a,m,J)}=\e^{-2\lambda_a(\varphi)}
*\left(d\Phi^{(a,m,J)}\wedge\tau(J)\right),
\ear
$a\in\tri$, $I\in\Omega_{a,e}$, $J\in\Omega_{a,m}$ and
$\Omega_{a,e},\Omega_{a,m}\subset \Omega_0$.
In (\ref{2.29n}) $*=*[g]$ is the Hodge operator on $(M,g)$.

For the potentials in (\ref{2.28n}), (\ref{2.29n}) we put
\beq{2.28nn}
\Phi^s=\Phi^s(x),
\eeq
$s\in S$, where
\beq{6.39i}
S=S_e\sqcup S_m,  \qquad
S_v\equiv \coprod_{a\in\tri}\{a\}\times\{v\}\times\Omega_{a,v},
\eeq
$v=e,m$.

For dilatonic scalar fields we put
\beq{2.30n}
\varphi^\alpha=\varphi^\alpha(x),
\eeq
$\alpha=1,\dots,l$.

>From  (\ref{2.28n})  and (\ref{2.29n}) we obtain 
the relations between dimensions of $p$-brane 
worldsheets and ranks of forms: 
$d(I) = n_a - 1,  \quad I \in \Omega_{a,e}$;
$d(J) = D - n_a - 1,  \quad J \in \Omega_{a,m}$,
in electric and magnetic cases respectively.

\section{Sigma model representation}
\setcounter{equation}{0}

We consider the case $d_0 = {\rm dim } M_0 \neq 2$ and use
generalized  harmonic gauge
\beq{3.1}
\gamma = {\gamma}_{0}(\phi) =
\frac{1}{2- d_0}  \sum_{i =1}^{n} d_i \phi^i
\eeq
($d_i = {\rm dim } M_i$).
For the model under consideration
the equations of motion 
and Bianchi identities $dF^s = 0, s \in S$,  are equivalent
to the equations of motion for the $\sigma$-model
with  certain constraints  imposed \cite{IMC}.

The $\sigma$-model action reads \cite{IMC}
\beq{3.2}
S_{\sigma} =
     \int_{M_0} d^{d_0}x \sq {g^0 }
\Bigl\{
{R}[g^0]
- G_{ij} g^{0\ \mu \nu} \p_{\mu} \phi^i  \p_{\nu} \phi^j -
2 {V}(\phi) - {\cal L}   \Bigr\},
\eeq
where
\beq{3.3}
G_{ij} = d_i \delta_{ij} + \frac{d_i d_j}{d_0 -2}
\eeq
are the components of the ("purely gravitational")
midisuperspace  metric on
$\R^{n}$  \cite{IM1}, $i, j = 1, \ldots, n$,
and
\beq{3.4}
 V = {V}(\phi)
= \Lambda e^{2 {\gamma_0}(\phi)}
-\frac{1}{2}   \sum_{i =1}^{n} \xi_i d_i e^{-2 \phi^i
+ 2 {\gamma_0}(\phi)}
\eeq
is the potential and 
\bear{2.30a}
{\cal L} = 
h_{\alpha\beta}
g^{0\ \mu \nu} \p_{\mu} \varphi^\alpha   \p_{\nu} \varphi^{\beta}
+
\sum_{a \in \Delta} \sum_{s  \in S}
\eps_s \exp(-2 U^s) 
g^{0\ \mu \nu} \p_{\mu} \Phi^{s}  \p_{\nu} \Phi^{s}.
\ear
Here 
\bear{2.u}
 U^s = U^s(\phi,\varphi)= -\chi_s\lambda_{a_s}(\varphi) +
\sum_{i \in I_s}d_i\phi^i, \\ \label{2.e}
\eps_s=(-\eps[g])^{(1-\chi_s)/2}\eps(I_s)\theta_{a_s}
\ear
for $s=(a_s,v_s,I_s)\in S$, $\eps[g]= \sign \det (g_{MN})$,
and  $\chi_s=+1$, for $v_s=e$, and $\chi_s=-1$, for $\quad v_s= m$.

\section{Exact solutions}
\setcounter{equation}{0}

In \cite{IMC,IMR} the Majumdar-Papapetrou type solutions were obtained 
in orthogonal case $(U^s,U^{s'}) = 0, s \neq s'$.
(For non-composite case see \cite{IM1,IM2}). These solutions correspond 
to Ricci-flat $(M_i,g^i)$, $i=0,\dots,n$, and were generalized also to 
the case of Einstein internal spaces \cite{IMC}. In \cite{IM} the 
Toda-lattice generalization of the "orthogonal" intersection rules was 
obtained and cosmological and spherically symmetric 
(classical and quntum)  solutions were considered.

\end{document}